# Cycle Accurate Binary Translation for Simulation Acceleration in Rapid Prototyping of SoCs[*]

Jürgen Schnerr[1], Oliver Bringmann[1], and Wolfgang Rosenstiel[1,2]

[1] FZI Forschungszentrum Informatik
Haid-und-Neu-Str. 10–14
76131 Karlsruhe, Germany

[2] Universität Tübingen
Sand 13
72076 Tübingen, Germany

{jschnerr, bringmann, rosenstiel}@fzi.de

**Abstract**

*In this paper, the application of a cycle accurate binary translator for rapid prototyping of SoCs will be presented. This translator generates code to run on a rapid prototyping system consisting of a VLIW processor and FPGAs. The generated code is annotated with information that triggers cycle generation for the hardware in parallel to the execution of the translated program. The VLIW processor executes the translated program whereas the FPGAs contain the hardware for the parallel cycle generation and the bus interface that adapts the bus of the VLIW processor to the SoC bus of the emulated processor core.*

## 1. Introduction

Nowadays, semiconductor chips reach such high complexities, that the integration of complete systems with high functionality on a single chip (so called Systems-on-Chip, SoC) becomes possible. Using an SoC bus, these SoC components are connected on the chip.

The development of software for such SoCs is problematic as the software running on SoCs usually includes very hardware near parts like device drivers. The requirements for this software are that I/O accesses to the bus must be cycle accurate in order to make it possible to validate the bus interfaces to the hardware or the handshakes on the bus.

Existing solutions for this problem are the hardware/software co-simulation [11] of processors and the attached hardware on the bus or the emulation of the processor using a hardware emulator. Using hardware/software co-simulation, the most accurate way of simulation would be a simulation of the processor and the attached hardware on an HDL simulator. But this leads to very slow execution times. To speed up the simulation, an abstraction of the processor has to take place. This can usually be done in two different ways. Either the source code has to be recompiled for the processor of the machine the simulation is running on, or an *instruction set simulation* (ISS) has to be used. An ISS can be implemented as an interpreted ISS, a just-in-time compiled ISS, or as a compiled ISS. Using a *bus functional model* (BFM), the recompiled code or the ISS has to be coupled to an HDL simulator simulating the hardware attached to the processor.

These existing solutions using hardware/software co-simulation suffer from different disadvantages, especially concerning accuracy and/or execution speed. Further disadvantages are, that some solutions require the source code of the program. Also if the hardware has to be simulated, a model of the hardware environment is needed.

Using hardware emulators or FPGAs for the emulation of the processor can lead to a significant speed-up of execution speed [12]. The main advantage of such solutions is that a very accurate emulation of the processor is possible at a reasonable speed. The disadvantage is that the detail level of the emulation cannot be chosen, therefore it can be more detailed than needed. Also a synthesizable specification of the processor is needed.

The rapid prototyping system used in this paper was presented in [13]. It is based on the hardware presented in [5, 6]. This system was designed to overcome the disadvantages of previous existing systems by offering a fast execution speed of the simulated system with sufficient cycle accuracy. To find a trade-off between execution speed and accuracy, several detail levels of code execution are offered. Another advantage compared to other solutions is, that neither a HDL description of the processor nor source code of the running program is needed.

Basically, the presented prototyping system consists of two parts. One part is the TMS320C6201 (C6x) VLIW processor from Texas Instruments [14], the other part consists of FPGAs. The VLIW processor is used for the emulation of the instruction set of the SoC processor core. As examples of processor cores the TriCore processor from Infineon [7] and the ARM processor [1] were used.

This paper will focus on the emulation of the processor core, that we have implemented using a compiled ISS. The binary translator that carries out the compiled ISS will be described in more detail in this paper. Compared to existing binary translators, this one annotates the translated program with instructions that access the cycle generation hardware in the FPGAs. The FPGAs of the system are also used for a bus interface between the C6x processor and the attached hardware that expects to be connected to an SoC bus.

Basically, the cycle accurate translation works as follows: The runtime of each basic block of the untranslated program is predicted. At the beginning of the translated basic block, code is included that triggers the start of the cycle generation with the number of predicted cycles. When such a translated basic block is run, it starts the cycle generation with its first instruction, and then the rest of the basic block runs in parallel to the cycle generation. At the end of the basic block there is an instruction that synchronizes the cycle generation with the instruction execution before the execution of the next basic block begins.

---

[*] This work has been partially supported by the BMBF/MEDEA+ project SpeAC A508 under grant number 01M 3067 A.

This method of letting the translated program trigger the clock cycles for the hardware itself reduces the bottleneck between the processor executing the program and the attached hardware.

The remainder of this paper is structured as follows: Section 2 shows a short overview on the state of art in using recompiled source code and ISS under consideration of timing issues. Section 3 of this paper describes the implementation of the cycle accurate binary translator. Section 4 provides results. Section 5 concludes the paper and gives an outlook.

## 2. State of the art

As stated in the previous section, a recompilation of the source code or an ISS can be coupled to a BFM to provide a cycle accurate abstraction of a processor.

Using a recompilation of the source for the target processor usually changes the timing of the program completely. To include the timing of the original processor there are several possibilities. One possibility is to add timing information in the form of delay annotation to the source code [2, 15]. Another approach to the inclusion of timing information was shown in [8]. In this case, the authors modified a C compiler to analyze the program being translated and to include the timing information into the generated code.

When a recompilation of the source code is not possible or too inaccurate, an ISS can be used. There are three possibilities for the implementation of such an ISS. The first one is the interpretative simulation. At the moment, this is the most commonly used method. As the interpretative simulation has to decode the instructions during the runtime, it suffers from low performance. To overcome this problem, a just-in-time (JIT) compilation (also called dynamic compilation) of the program can be used [10]. Here the performance is much better as the program is translated during the runtime and blocks of the translated program are cached leading to a much better performance if the same part of the program has to be executed more than once.

The third possibility is the compiled simulation (also called binary translation or static compilation) of the program. As the decoding and the translation of the program are done statically during the compile-time it can reach the fastest execution speed of the three possible implementations.

The development of binary translators started at the end of the 1980s where they were needed for the translation of existing CISC programs in order to migrate to RISC architectures. For this reason they offered only a functional translation of the program with no consideration of timing issues. A good introduction and an overview of such translators are shown in [3].

For the translation of programs that access hardware these purely functional solutions were not enough. Therefore, Cogswell and Segall presented in [4] a methodology of translating the program where the translated program is instrumented with timing information in a way that during the execution of the program on the target platform the timing for the source platform contained in the timing information and the timing on the target platform can be compared at regular intervals and the program execution can be delayed if it is necessary.

This paper will introduce a cycle accurate binary translation specially tailored to the needs of prototyping of SoCs. In contrary to previous solutions, it works on a very low, hardware near level.

## 3. Cycle accurate static compilation

This section will describe the compiler presented in this paper in more detail. The design of the compiler is made up of two parts. One part consists of the description of the instruction set and the architecture of the processor whose code has to be translated (the source processor). This description is done using C++ classes. The other part is independent of the source processor and is implemented as a library. To generate a working compiler both parts have to be linked together. This way, the compiler can be adapted to different source processors.

Although it is possible to write the classes containing the description of the source processor by hand, it is not recommended due to its error-proneness. Therefore, this processor is usually defined in an XML file that is translated into the appropriate C++ code by a tool. This XML file contains an architecture description and a description of the instruction set of the processor. The architecture description contains a description of the pipelines and the caches of the processor, whereas the instruction set description contains information about instruction decoding and the semantics of the described instruction written in an intermediate code which resembles the assembler instructions of the C6x processor but does not have their constraints.

Figure 1 shows an overview of how the compiler works. The parts of the compiler needed for cycle accurate translation of code are shown in grey.

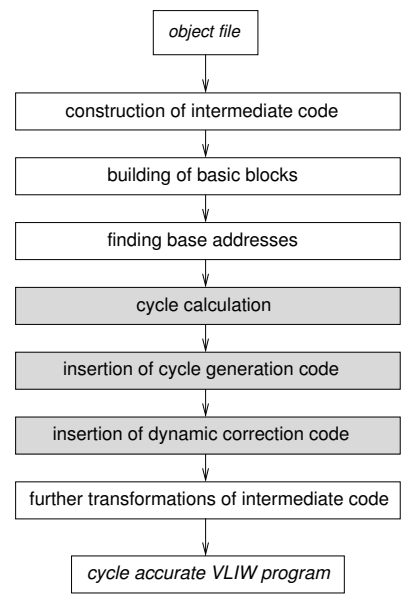

**Figure 1. Cycle accurate static compiler**

Using an appropriate class, the compiler reads the object file, which is usually provided in ELF format (other formats are also possible). Afterwards, this object code will be decoded and translated into an intermediate representation. This intermediate representation consists of a list of objects. Each of these objects is representing one intermediate instruction.

In a next step, the basic blocks of this program are found out using the list containing the translated program, and a list of basic blocks is built.

Using the list of basic blocks, the base addresses of load/store instructions have to be found out, as far as this is statically possible. This is needed for two reasons. One reason is to change the base addresses of load/store instructions accessing memory to the new memory addresses of the target system. The other reason is to find out, which of these load/store instructions are I/O instructions. These I/O instructions have to be replaced by instructions accessing the hardware of the bus model.

After that, a static calculation of the number of cycles each basic block would have taken on the source processor is made. How this calculation works is described in more detail in Section 3.3.

In the next step, the code for the cycle generation for each basic block has to be inserted into the basic blocks. How this code looks like and how it works is described in Section 3.1.

Not every effect of the processor architecture on the number of cycles can be predicted statically. Therefore, if effects of branch prediction and caches should be taken into account, additional code has to be added in this step. Further details concerning this code are described in Section 3.4.

After that, further transformations of the intermediate code have to be done. Here, instructions that can be executed in parallel on the VLIW processor have to be found out. Furthermore, software pipelining of the instructions and register binding have to take place. Also, every instruction has to be assigned to the functional unit it will run on.

The result is a program for the TMS320C6x VLIW processor annotated with cycle information. More information about how this annotation works is described in the next section.

### 3.1. Annotation of translated code

Usually, there is an external common clock for the processor and the attached hardware, but in this case there is no such common clock. Instead, the emulated processor has to generate the clock cycles for the attached hardware itself.

This is done using accesses to a hardware device in the FPGAs called synchronization device.

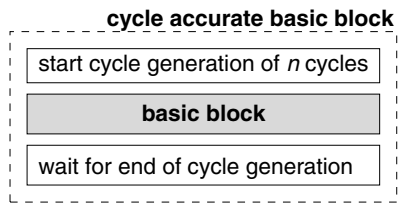

**Figure 2. Annotated basic block**

As shown in Figure 2 the compiler adds an instruction that starts the cycle generation at the beginning of the basic block. This instruction is a write access to the synchronization device that contains the number $n$ of cycles this basic block would need on the source processor. From now on the execution of the instructions in the translated basic block and the generation of the cycles for the attached hardware run in parallel until the executed program reaches the "wait for end of cycle generation" instruction. This instruction consists of a read access to the synchronization device. If the cycle generation is already finished, this instruction does nothing. If it is not finished, this instruction waits until the generation is finished.

This way, the execution of the translated program and the attached hardware can be synchronized without having the bottleneck of permanent hardware accesses by the executed program.

### 3.2. Cycle calculation

In order to guarantee an as fast as possible execution of the translated code, static cycle calculation, as mentioned in the previous section, should be used if it is possible. But compared to a dynamic cycle calculation this leads to less accuracy as not every effect of processor architecture on the number of executed cycles can be exactly predicted.

In order to guarantee both an as fast as possible execution of the code as well as the highest possible accuracy, it is possible to choose different levels of accuracy of the generated code when the compiler is run.

There are the following three detail levels of cycle accuracy:

1. purely static prediction

2. dynamic improvement of the static prediction
   (modeling of the branch prediction)

3. dynamic inclusion of instruction caches
   (additionally to point 2)

The cycle calculation in these different detail levels will be discussed in more detail in the following text.

### 3.3. Static cycle calculation of a basic block

The simplest possible way of calculating the execution time a basic block consists of the summation of the execution or the latency times of the instructions of this basic block. But this is only possible with simple architectures. Concerning modern architectures, it is too inaccurate as it does not consider pipeline effects, super scalarity and caches of these architectures.

In order to predict pipeline effects and the effects of super scalarity statically, modeling the pipeline per basic block becomes necessary [9]. Using information about the instructions and the pipelines of the processors, it can be found out which instructions of the basic block can be executed in parallel with a super scalar processor and which combinations of instructions in the basic block cause pipeline stalls.

With the information gained by the modeling of the basic block a prediction can be made how many cycles the basic block would have needed on the source processor.

In the following section, it will be shown how such a prediction can be improved during runtime.

### 3.4. Dynamic correction of cycle prediction

Not every cycle that a processor executes can be correctly predicted in a static way. For example, if a conditional branch occurs at the end of a basic block or the effects of instruction caches during the execution of a basic block.

In these cases, additional code that counts correction cycles has to be inserted into the translated basic block, and at the end of the basic block a correction block has to be inserted which generates these additional correction cycles before the next basic block can be run.

Figure 3 shows such an annotated basic block that offers additional dynamic correction of the cycle count. The additional cycle counting code for the consideration of branch prediction is explained in the next section. The division of the basic block for the calculation of additional cycles for instruction cache misses as shown in the Figure 3 is explained in Section 3.4.2.

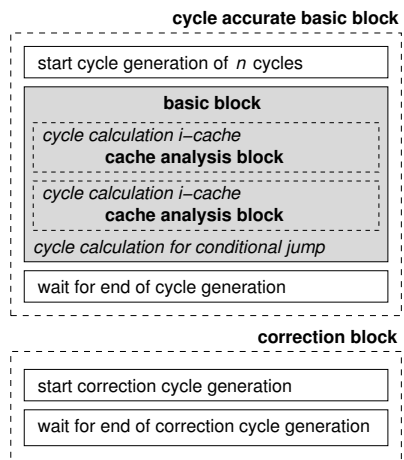

**Figure 3. Annotated basic block with dynamic correction**

**3.4.1. Branch prediction** Conditional branches usually have different cycle times considering whether the branch prediction was right or wrong and whether the branch was taken or was not taken. Usually, such a conditional branch needs a minimum number of cycles in all cases. This number can be added to the cycle generation code of the basic block.

Whether there are additional cycles needed or not, has to be found out using additional code which has to be inserted in this step just before the conditional branch. During runtime, the inserted code has to check whether the branch prediction was correct or not and whether the branch is taken or not. The additional cycles will be added to a cycle counter for the execution in the correction block.

**3.4.2. Instruction cache** For additional simulation of the instruction cache, three steps have to be done. First, space has to be reserved at the end of the translated program. This space is used to hold the data of the simulated cache during runtime. The second step that has to be done, is that every basic block of the translated program has to be divided into cache analysis blocks. In a third step, code for the cache handling has to be added to each of these cache analysis blocks.

All three steps will be explained in detail in the following three sections.

**Saving cache data** At the end of the translated program space for cache data is added. This space holds the valid bit, the cache tag and the least recently used (*lru*) information (containing the replacement strategy) for each cache set during the runtime. To simplify the handling of the cache tag and the valid bit, they are combined into one word which is usually 32 bits long.

The number of cache tags and the according valid bits is depending on the associativity of the cache, e.g. for a two way set associative cache there are two of them.

**Cache analysis blocks** For the consideration of the effects of instruction caches, each basic block has to be divided into cache analysis blocks. Each cache analysis block contains that part of a basic block that fits into a single cache block. Every instruction of such a cache analysis block has the same tag and the same cache index.

To find out the cache analysis blocks of a basic block all addresses of the untranslated instructions of this basic block have to be examined. Each address consists of tag information, cache index and block offset information.

The cache index information determines at what position of the cache the instruction with this address is cached. The tag information has to be used to find out which address had been cached, as there can be multiple addresses with the same cache index. The block offset is the position in the cache line at which the instruction is cached. For this simulation, the block offset information is not needed, as no real caching of data takes place.

Now that the cache analysis blocks have been found out, code that handles the simulated cache and calculates the additional cycles of cache misses has to be added to these blocks. How this is done is described in the next section.

**Cycle calculation code** During runtime, it has to be found out, for each cache analysis block the basic block consists of, whether it is in the simulated cache or not. This way, cache misses can be detected.

Each cache analysis block is characterized by a tag and cache set index combination. At the beginning of each cache analysis block there is a branch to a subroutine included. This subroutine, shown in Figure 4, gets the tag and cache set combination as parameters. It checks whether the tag of the cache analysis block can be found in the specified set and whether the valid bit for the found tag is set.

---

**input:** tag and cache set index of cache analysis block
**output:** changed cache information and cycle correction
       counter

**if** *tag can be found in specified set* **and** *valid bit is set* **then**
  *renew least recently used information*  // cache hit
**else** // cache miss
  *use lru information to find out tag to overwrite*
  *write new tag*
  *set valid bit of written tag*
  *renew lru information*
  *add additional cycles to cycle correction counter*
**return** *to normal execution of the cache analysis block*

---

**Figure 4. Cycle correction for caches**

If yes, the block is already cached und no additional cycles are needed. Only the *lru* information has to be renewed.

If no, the *lru* information has to be used to find out which tag has to be overwritten. After that, the new tag has to be written instead of the found old one, and the valid bit for this tag has to be set. Also the *lru* information has to be renewed. In a last step, the additional cycles have to be added to the cycle correction counter.

After that, the subroutine returns, and the normal execution of the instructions of the cache analysis block can continue.

The code of the subroutine is automatically generated by the translator on basis of the description it has about the cache of the source processor. It is appended to the translated code. In large basic blocks, this code can be included into the basic block making the subroutine call unnecessary and the parallel execution of the cache calculation code and the executed program on the VLIW processor possible.

### 3.5. Debugging of translated code

The debugging of code annotated with cycle information is implemented using an interface program between the translated code and the remote debugging interface of the GNU Debugger (gdb). One important problem which had to be solved was the implementation of a single step possibility for the debugger as the translated code only contains cycle generation for complete basic blocks.

This has been solved in the way that the debug code contains two translations of the original code. In one of these translations the code has to be annotated with a basic block oriented cycle generation, and in the other one it has to be annotated with an instruction oriented cycle generation. The code with the instruction oriented generation of cycles contains the cycle generation for each translated instruction and a branch after each such instruction into the ROM routines that communicate with the debug interface.

Using the described two translations of the code, the debug interface has to implement break points, single step execution and normal program execution. Also using information made during the translation of the program, the debug interface has to translate the register names and the addresses used.

Break points that occur in a certain basic block are always set at the beginning of a basic block of the basic block oriented translated code. To get to the real break point the single step program has to be used.

## 4. Results

In order to test the execution speed and the accuracy of the translated code, a few examples were compiled using a C compiler into TriCore object code. This object code was then translated into cycle accurate TMS320C6x binary code and executed on the emulation platform. As a reference, the execution speed and the cycle count of the TriCore code has been measured on a TriCore TC10GP evaluation board, using the methodology described in [16].

The examples consist of two more control flow dominated programs (gcd, sieve), two filters (fir, ellip), and two programs that are part of audio decoding routines (dpcm, subband).

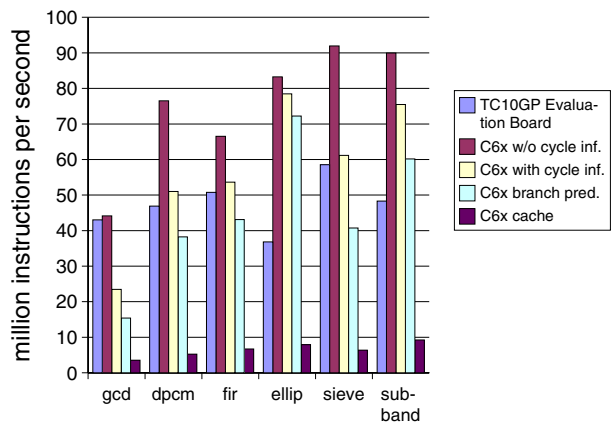

**Figure 5. Comparison of speed**

The comparison of the execution speed of the generated code with the execution speed of the TriCore evaluation board is shown in Figure 5. The C6x processor on the emulation system ran at a clock rate of 200 MHz. The TriCore processor of the evaluation board ran at 48 MHz, which is about one fourth of the clock rate of the C6x processor.

A fast execution speed of the cycle accurate code can be gained especially for examples with large basic blocks like ellip and subband. This is due to the fact that a program containing larger basic blocks needs fewer cycle generation instructions and fewer branches. Branches are very costly in regard to the cycle time on the C6x processor. Also larger basic blocks allow a better parallelization of the instructions they contain on the VLIW processor, leading to a smaller execution time.

On the other side, the program sieve contains many small basic blocks. Each one of these basic blocks has its own cycle generation code. Therefore, the translated code containing the cycle information is much slower than the translated code without information as can be seen in Figure 5.

| TC10GP Evaluation Board | 1.08 |
| --- | --- |
| C6x without cycle information | 2.94 |
| C6x with cycle information | 4.28 |
| C6x branch prediction | 5.87 |
| C6x caches | 35.34 |

**Table 1. Clock cycles per TriCore instruction**

The average number of clock cycles needed for the execution of one TriCore instruction is shown in Table 1. The results on the C6x processor are the average value of all examples.

For the compiled code without cycle information the C6x processor needs about three cycles to execute a translated TriCore instruction. This relatively low overhead results from the fact that in the translated code of these examples on the average about two or three C6x instructions can be executed in parallel.

The addition of cycle information to the code adds on the average only a little more than one clock cycle per translated TriCore instruction. Also the overhead for the additional TriCore branch prediction is very low.

On the next detail level, the additional consideration of instruction caches, about six times more cycles are needed in comparison to the code containing the branch prediction. This is due to the fact that for each cache analysis block a complete routine has to be run to find out whether there is a cache miss or a cache hit.

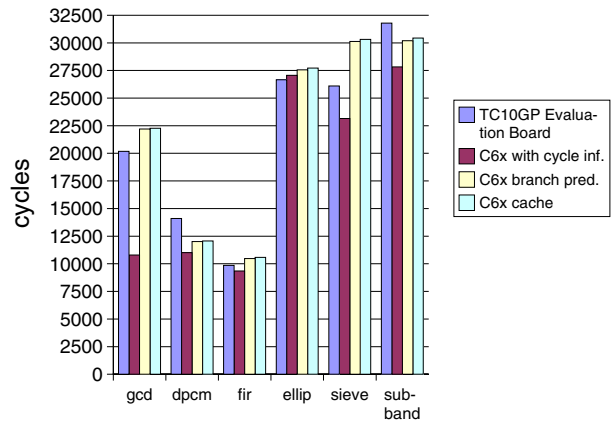

**Figure 6. Comparison of cycle accuracy**

A comparison of the number of simulated cycles of the generated code in different detail levels with the number of executed cycles of the TriCore evaluation board is shown in Figure 6. The deviation of the cycle counts of the translated programs (with branch prediction) compared to the measured cycle count ranges between 3 % for the program ellip to 15 % for the program sieve. Especially for control flow oriented programs like gcd it can be seen that branch prediction can play an important role.

|  | gcd | fibonacci | sieve |
|---|---|---|---|
| # of executed instructions | 1484 | 41419 | 20779 |
| **Simulation** (Workstation) | approx. 28 sec | approx. 10 min | approx. 18 min |
| **Emulation** (FPGA) | 321 μsec | 3.9 msec | 21.8 msec |
| **Translation** C6x cycle | 63.1 μsec | 950 μsec | 520 μsec |
| C6x branch | 94.6 μsec | 1.4 msec | 781 μsec |
| C6x cache | 416 μsec | 6.3 msec | 5 msec |

**Table 2. Software runtime comparison**

Table 2 provides a comparison with the results published in [12]. It uses the same three example programs and compares the translation results with the results running the programs on an RT level simulation of the TriCore processor core on a workstation and the result of the emulation of the core on a prototyping platform using a Xilinx XCV2000E FPGA. The emulation on this FPGA is running with a frequency of 8 MHz.

The table shows that the translation results considering caches are about in the same range as the emulation results on the FPGA. Whereas the translation results of the other two detail levels are significantly (28 and 42 times) faster for the sieve example and about 3 to 5 times faster for gcd and fibonacci. The relatively slow execution of the latter two examples results from the very short basic blocks they consist of.

Apart from the possibility to gain a much higher performance by lowering the detail level, the translation approach has another important advantage compared to the emulation of the processor core with an FPGA. As processors get more and more complex and as there is a capacity gap between the ASIC and FPGA gate capacities, it will remain problematic to map state-of-the-art processors onto FPGAs. Using the translation approach, a more complex processor only results into a more complex architecture and instruction set description for the translator.

## 5. Conclusion and outlook

In this paper, we have shown how a microprocessor can be emulated cycle accurately on a prototyping platform using the annotation of translated code.

Especially pipelines and caches had to be considered. It has been shown, how cycle generation code for these can be generated and inserted. This inserted code combined with the appropriate hardware adds only a low overhead to the translated code and allows a fast execution. To find a trade-off between accuracy and execution speed, the cycle accuracy of the translated code can have several detail levels. In addition to the possibility to choose the detail level of accuracy, another important advantage of the described emulation system is that no RT level code of the emulated processor has to be available.

There are several possibilities to enhance the system to make it either faster or more accurate. Execution speed improvements can be achieved by the addition of the possibility of cycle generation for blocks which are larger than the basic blocks that are used now. Another speed improvement can be, to combine the translation of the code with a static cache analysis. This could speed up the translated code regarding the simulation of caches. To improve the accuracy of the translated code, additional consideration of data caches can be added. Furthermore, the accuracy of certain instructions could be improved. A consideration of data dependent behavior of certain instructions like multiplications and divisions can also help to improve the accuracy in certain cases. For example, on a processor that uses a Booth multiplier the delay of this multiplier depends on operand value. Also, memory access delay should be taken into account for load/store instructions. Such a delay can happen if, for instance, a cache miss happens and both the processor (via cache) and a DMA controller want to access the memory.

Another enhancement of the system that is especially important for the execution of embedded software and the execution of operating systems will be the consideration of interrupt handling and exceptions.